\documentclass{elsart}

\usepackage{amssymb}
\usepackage{graphicx}

\begin{document}

\begin{frontmatter}

% Title
  
  \title{Magnetization anomalies in the superconducting state of RuSr$_{2}$GdCu$_{2}$O$_{8}$ 
and the magnetic study of Sr$_{2}$GdRuO$_{6}$ }

% Authors and addresses
  
\author[1]{Thomas P. Papageorgiou},
\author[1]{Thomas Herrmannsd\"orfer}, 
\author[2]{Robert Dinnebier\thanksref{address}},
\thanks[address]{Present address: Max-Planck-Institute for Solid State Research, 
Heisenbergstrasse 1, D-70569 Stuttgart, Germany}
\author[1]{Timo Mai}, 
\author[1]{Tobias Ernst} 
\author[2]{Markus Wunschel}
\author[1]{and Hans F. Braun}
  
\address[1]{Physikalisches Institut, Universit\"at Bayreuth, D-95440 Bayreuth,
  Germany}
\address[2]{Lehrstuhl f\"ur
  Kristallographie, Universit\"at Bayreuth, D-95440 Bayreuth, Germany}

% Text of abstract

\begin{abstract}

d.c. magnetization measurements performed on a polycrystalline sample of 
\linebreak RuSr$_2$GdCu$_2$O$_8$ (Ru-1212) showed distinct peaks of the 
magnetization upon entering the superconducting state. Since Sr$_2$GdRuO$_6$ (Sr-2116) 
is the precursor for the preparation of Ru-1212, 
a detailed investigation of the magnetic properties of Sr-2116 was carried out. 
Although similarities were observed in the magnetic behavior 
of Sr-2116 and Ru-1212 in the temperature range of the observed peaks, 
we can exclude, based on a quantitative comparison, that the anomalies observed for 
Ru-1212 are due to Sr-2116 impurities.  

\end{abstract}

% Keywords and PACS


\end{frontmatter}

% Main text

\section{Introduction}
\label{sec:intro}

The discovery of coexistence of superconductivity and weak ferromagnetism in
RuSr$_2$GdCu$_2$O$_8$ (Ru-1212) \cite{Bauernfeind2,Bauernfeind3,Bauernfeind1} 
has triggered a large number
 of studies of the properties of this material. From all these
studies it becomes obvious, that the preparation method plays a very important
role, in particular as far as the superconducting properties of the compound are
concerned. For example, there are reports of non-superconducting samples of Ru-1212
\cite{Felner} as well as for samples, which exhibit bulk superconductivity
\cite{Bernhard1}. A key compound for the preparation of Ru-1212
is Sr$_2$GdRuO$_6$ (Sr-2116), either as an intermediate product during the
preparation of Ru-1212 \cite{Bauernfeind3,Bernhard2}, or used directly as a
precursor \cite{Bauernfeind1}. The (nominal) valency of Ru in Sr-2116
(Ru$^{5+}$) minimizes the likelihood of  formation of the itinerant
ferromagnet SrRuO$_{3}$
during the preparation of Ru-1212 in the two-step process \cite{Bauernfeind3}.

Sr$_2$GdRuO$_6$ belongs to the class of double perovskites of the general 
chemical formula A$_2$BB$^{'}$O$_6$ (often called ``2116'' compounds 
because of their stoichiometry) with A=Ca, Sr, Ba, B=Y, La, Nd,
Eu, Gd-Lu and B$^{'}$=Ru. Their structure, as determined by neutron 
\cite{Battle1,Battle2,Battle3,Battle4,Battle5} and X-ray diffraction
\cite{Doi}, is that of a distorted perovskite with monoclinic 
space group P2$_{1}$/n and can be visualized as corner-sharing tilted oxygen
octahedra alternately centered by  B and B$^{'}$, while the
bigger A ions are located in the area between the oxygen octahedra.

These compounds  offer the rare opportunity to study the magnetic behavior
of a 4d$^3$ electron system. Indeed, magnetic ordering of highly oxidized
cations from the second transition series is quite unusual but
Battle \textit{et al.} \cite{Battle1,Battle2,Battle3,Battle4,Battle5} 
have shown, that the Ru$^{5+}$ ions in these compounds order
antiferromagnetically at low temperatures.

A particular interest for a more detailed investigation of the magnetic
properties of Sr$_2$GdRuO$_6$ arises from the fact mentioned above, that this
compound is involved in the preparation methods of the magnetic superconductor
Ru-1212. In this work we investigate samples of Ru-1212 prepared using
Sr-2116 as a precursor material and examine the origin of
distinct peaklike anomalies in the d.c. magnetization curves of Ru-1212
upon entering the superconducting state. We conclude, that these peaks 
cannot be explained in terms of some of our precursor
material remaining as an impurity phase in the superconducting samples. 
A tentative explanation of the origin of the peaks, 
based on the interpretation of our data on Sr-2116, is given.

\section{Experimental}\label{sec:exp}

Polycrystalline samples of Ru-1212 were prepared following a two-step
procedure. First, Sr-2116 was prepared from stoichiometric quantities of
RuO$_2$, Gd$_2$O$_3$ and SrCO$_3$.  The powders were mixed, calcined at
950~$^{\circ}$C in air, pressed into pellets and fired for 16~h at
1250~$^{\circ}$C in air.  In a second step, the obtained Sr-2116 was mixed
with CuO, pressed into pellets and fired for 120~h at 1060~$^{\circ}$C in
flowing oxygen.

Resistance measurements were performed by a standard four-probe a.c. technique
(at 22.2 Hz) on bar-shaped pieces cut from the pellets using silver paint
contacts.  a.c. susceptibility measurements were done with a home-made
susceptometer using a standard lock-in technique at different field
amplitudes.

d.c. magnetization measurements were done with a commercial SQUID magnetometer
(Cryogenic Consultants Ltd. S-600), which allows measurements in the
temperature range 1.6~K $\leq$ \textit{T} $\leq$ 300~K in magnetic fields up
to 6 T. For the low field measurements, a paramagnetic \underline{Au}Cr sample was used as a standard
for the determination of the remanent fields in the magnet.
An external current source (Knick DC-Current-Calibrator J152) was used to
cancel the remanent and earth fields and apply the appropriate current, so that the
desired field could be achieved.

\begin{figure}
\begin{center}
  \hspace*{-4mm}\includegraphics[clip=true,height=48mm,width=75mm]{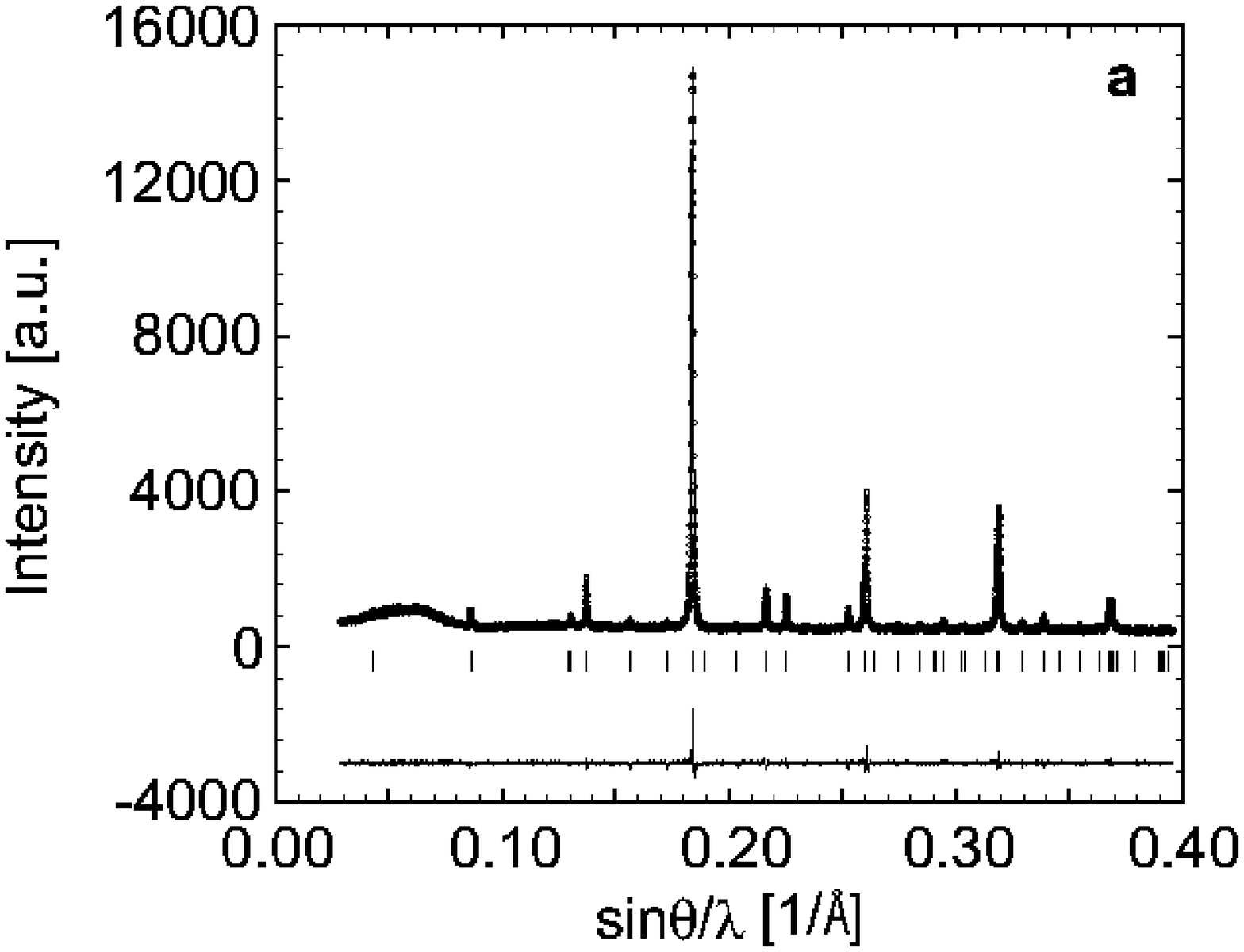}\\[2ex]
  \includegraphics[clip=true,width=75mm]{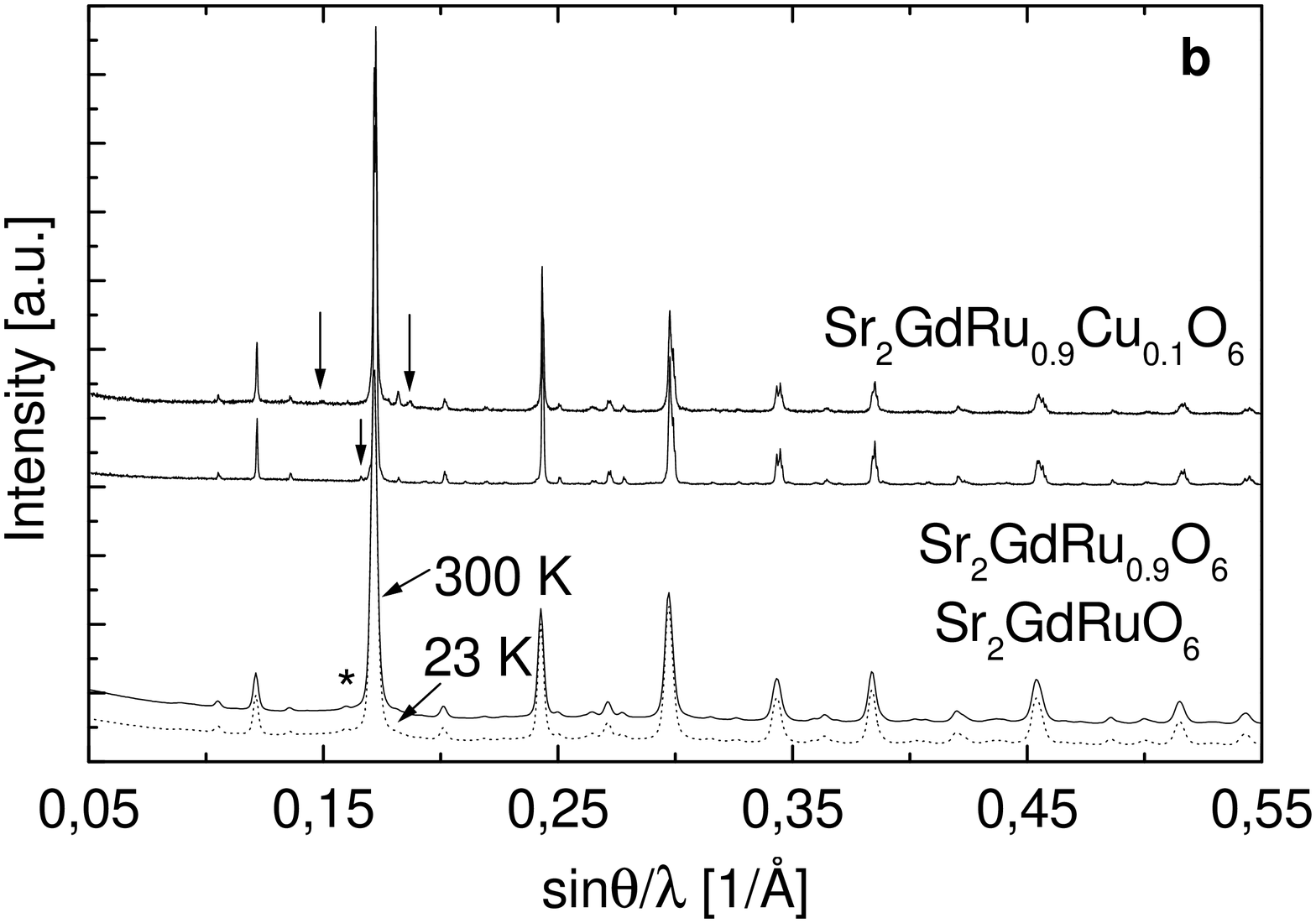}
\end{center}
\caption{(\textbf{a}) X-ray (CuK$\alpha$) diffraction spectrum of the Ru-1212
  sample. The solid line is the LeBail fit \cite{Lebail} in space group P4/mmm 
  done using the program FULLPROF \cite{Carvajal}. The peak positions
  and difference line are also shown. The amorphous hump at $\sim$ 0.05 in $\sin\theta/\lambda$ is due to the vacuum grease. 
  (\textbf{b}) X-ray patterns of Sr-2116
  and related compounds. The asterisk indicates the position of the strongest
  peak of Gd$_{2}$O$_{3}$ and the arrows the positions of unidentified impurity peaks.}\label{fig:xray}
\end{figure}

The Ru-1212 piece studied in the magnetometer was powdered afterwards and
the x-ray powder diffraction pattern was taken on a Philips X-pert
diffractometer with primary beam Ge(111) Johanson monochromator using Cu
K$\alpha$$_{1}$ radiation. Flat plate geometry was used with the sample 
sieved on vacuum grease on a low background quartz sample holder.
The result shown in figure~\ref{fig:xray}a indicates, that within the
resolution of the instrument the sample is single phase with the impurity
phases (if any) representing less than 3~\% of the sample. The
tetragonal lattice parameters (spacegroup P4/mmm) are
$a=3.837(4)~\AA, c=11.56(1)~\AA$.

A similar X-ray scan was done for the Sr-2116 sample studied in the
magnetometer with a Seifert XRD~3000~P diffractometer and revealed, that traces
of the original Gd$_2$O$_3$ powder were still present in the sample.  
The room temperature lattice parameters for our Sr-2116 sample were
(spacegroup P2$_{1}$/n) $a=5.808(5)~\AA, b=5.819(5)~\AA, c=8.212(8)~\AA, \beta=90.3(1)^\circ$.

X-ray
powder diffraction data of a similarly prepared Sr-2116 sample at low
temperatures were collected in transmission geometry 
at the high energy beamline ID15B of the 
European Synchrotron Radiation Facility
(ESRF).  X-rays of energy 90 keV were selected by a bent Si (511)
monochromator, which focuses the beam in the horizontal plane \cite{Suortii}.
The size of the beam was adjusted to several mm height and a width of 0.5 mm.
A MAR~345 image plate reader was used as detector at a distance of approx.
730~mm from the sample. The wavelength was determined to 0.14668~\AA\ from a
NBS-676 LaB$_{6}$ standard. 
The sample was contained in a 0.7 mm lithiumborate
glass (glass No. 50) capillary and was rocked several degrees during the
measurement in order to improve randomization of the crystallites. Data were
taken at several temperatures down to 23~K. 
Data reduction was performed using the program FIT2D
\cite{Hammersley1}, resulting in diagrams of corrected intensities versus the
scattering angle 2$\theta$. It was observed, that the diffracted intensity was
quite uniformly distributed over the Debye-Scherrer rings, ruling out severe
grain size effects and preferred orientation. Low angle diffraction peaks had
a FWHM of 0.04$^{\circ}$ 2$\theta$, which can be considered the highest
resolution for this particular setup. At this high energy, no corrections for
absorption are necessary. 
The data presented in the lower part of
figure~\ref{fig:xray}b show, that the crystal structure of the
compound is stable at least down to 23~K.
% Begin figure1
% End figure1
\section{Results and Discussion}\label{sec:results}

\subsection{Superconductivity and magnetism of RuSr$_2$GdCu$_2$O$_8$}\label{sec:resa}

\begin{figure}
\begin{center}
  \includegraphics[clip=true,width=75mm]{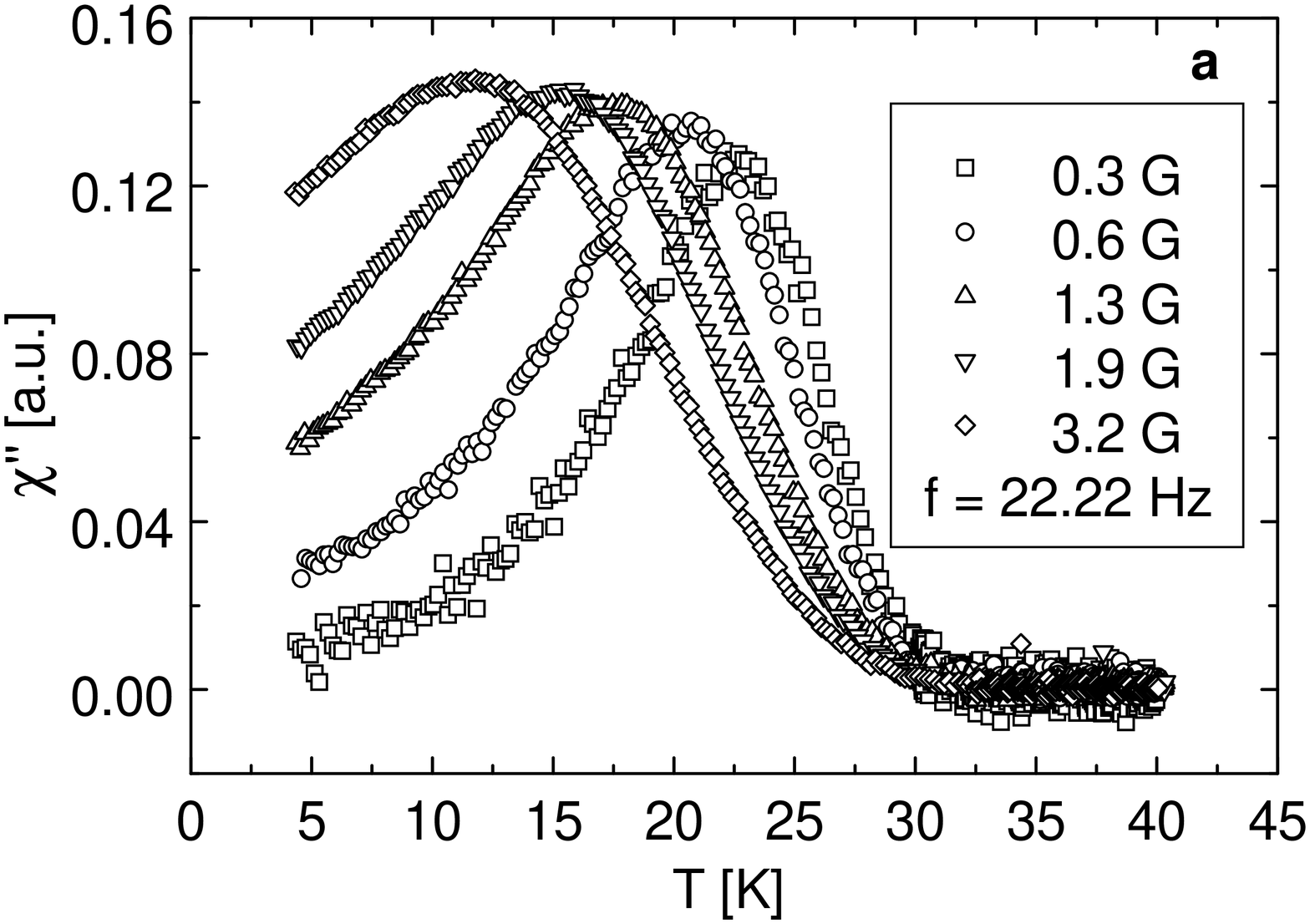}\\[2ex]
  \includegraphics[clip=true,width=75mm]{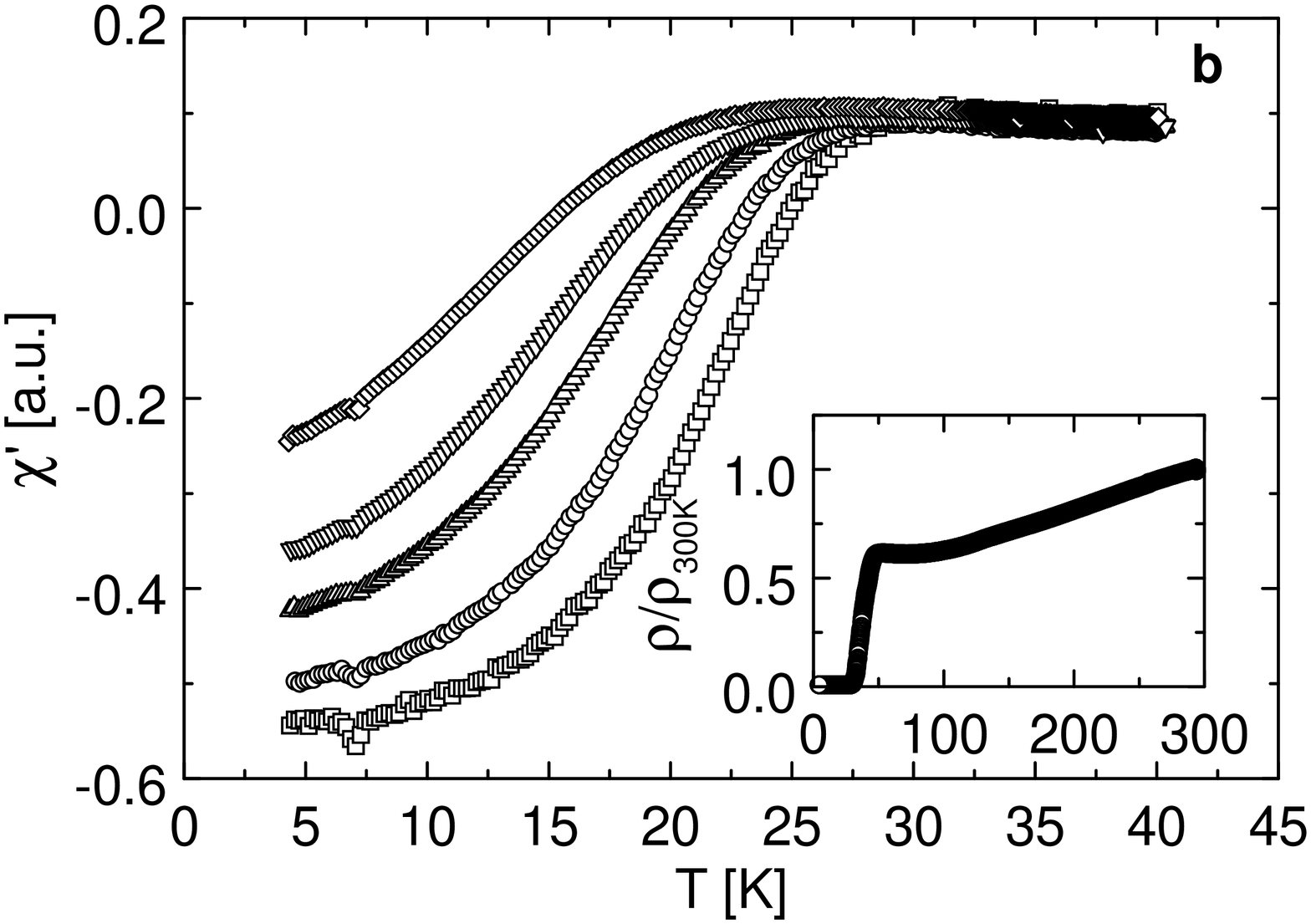}
\end{center}
\caption{The imaginary (\textbf{a}) and real (\textbf{b}) part of the a.c. susceptibility for Ru-1212 at low temperatures and different magnetic field 
  amplitudes. Inset: resistivity data normalized to the room temperature
  value.}\label{fig:res}
\end{figure}

In figure~\ref{fig:res} our resistivity and a.c. susceptibility results on Ru-1212 are
shown.
% Begin figure2
% End figure2
The resistivity measurements (inset) reveal a metallic behavior of the sample at
high temperatures with a small cusp at about 135~K related to the magnetic
transition, which occurs at this temperature (see d.c. magnetization
measurements presented below). At lower temperatures a slight upturn of the
resistance is observed above the onset of superconductivity at 50~K, while the
resistivity becomes zero at 30~K. At this temperature, the inter-granular
coupling is established and a clear diamagnetic response is observed in the
real part of the a.c. susceptibility with the corresponding loss peaks in the
imaginary part. 
Typical for shielding due to inter-granular coupling, the transition widens and
shifts to lower temperatures as the a.c.  field amplitude is increased.

\begin{figure}
\begin{center}
  \includegraphics[clip=true,width=75mm]{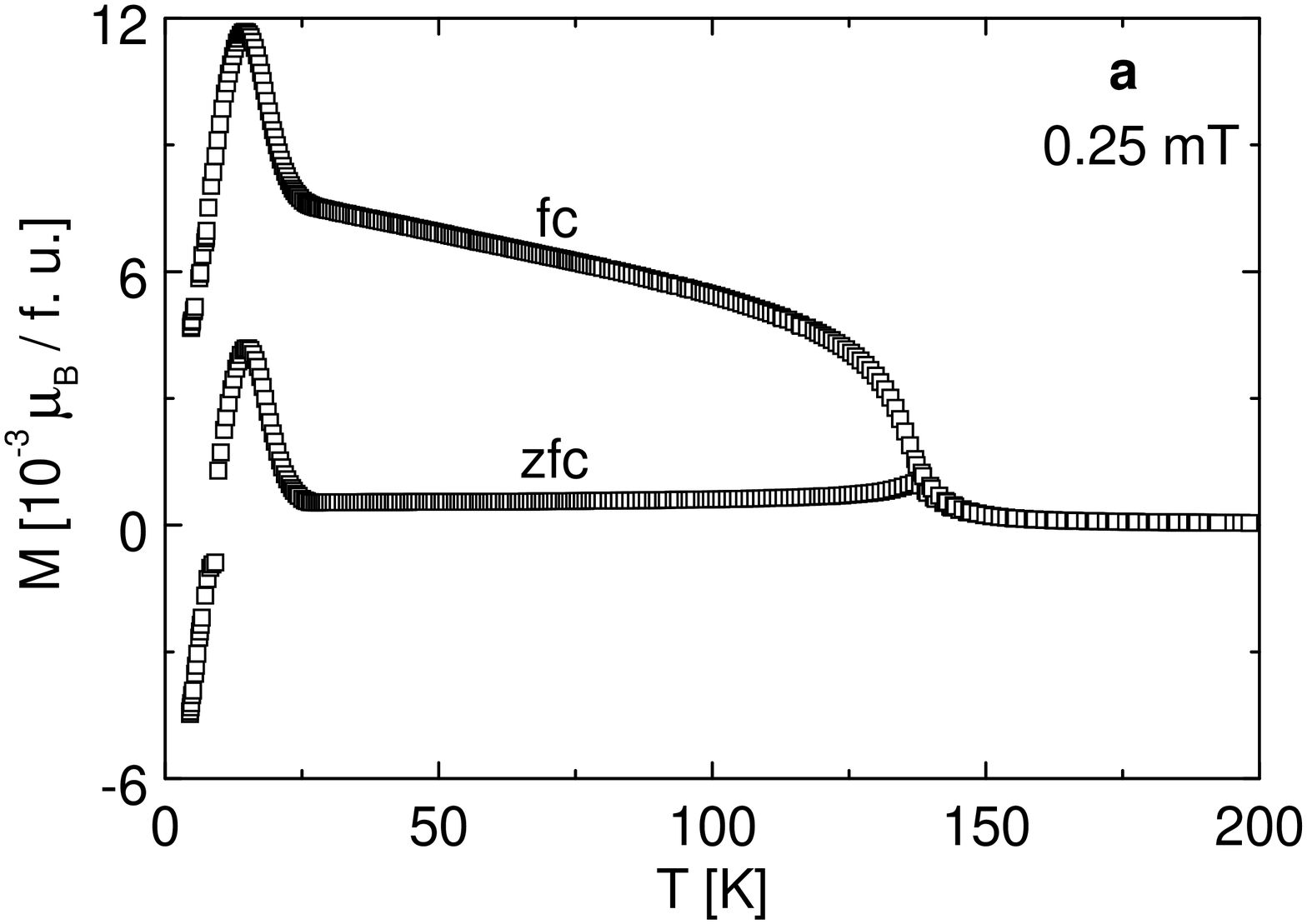}\\[2ex] %\vspace{-1cm}
  \includegraphics[clip=true,width=75mm]{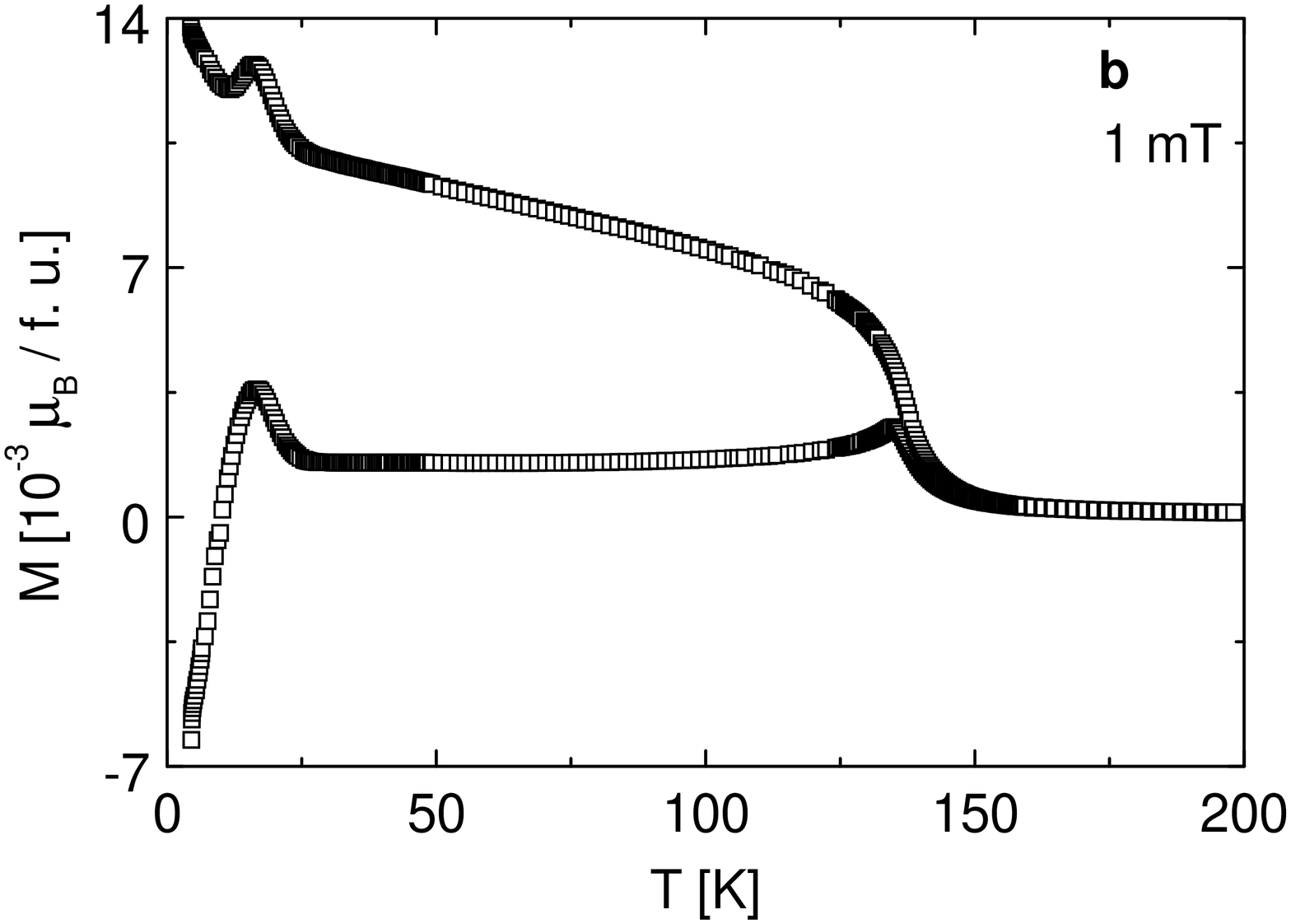}\\[2ex]
  \includegraphics[clip=true,width=75mm]{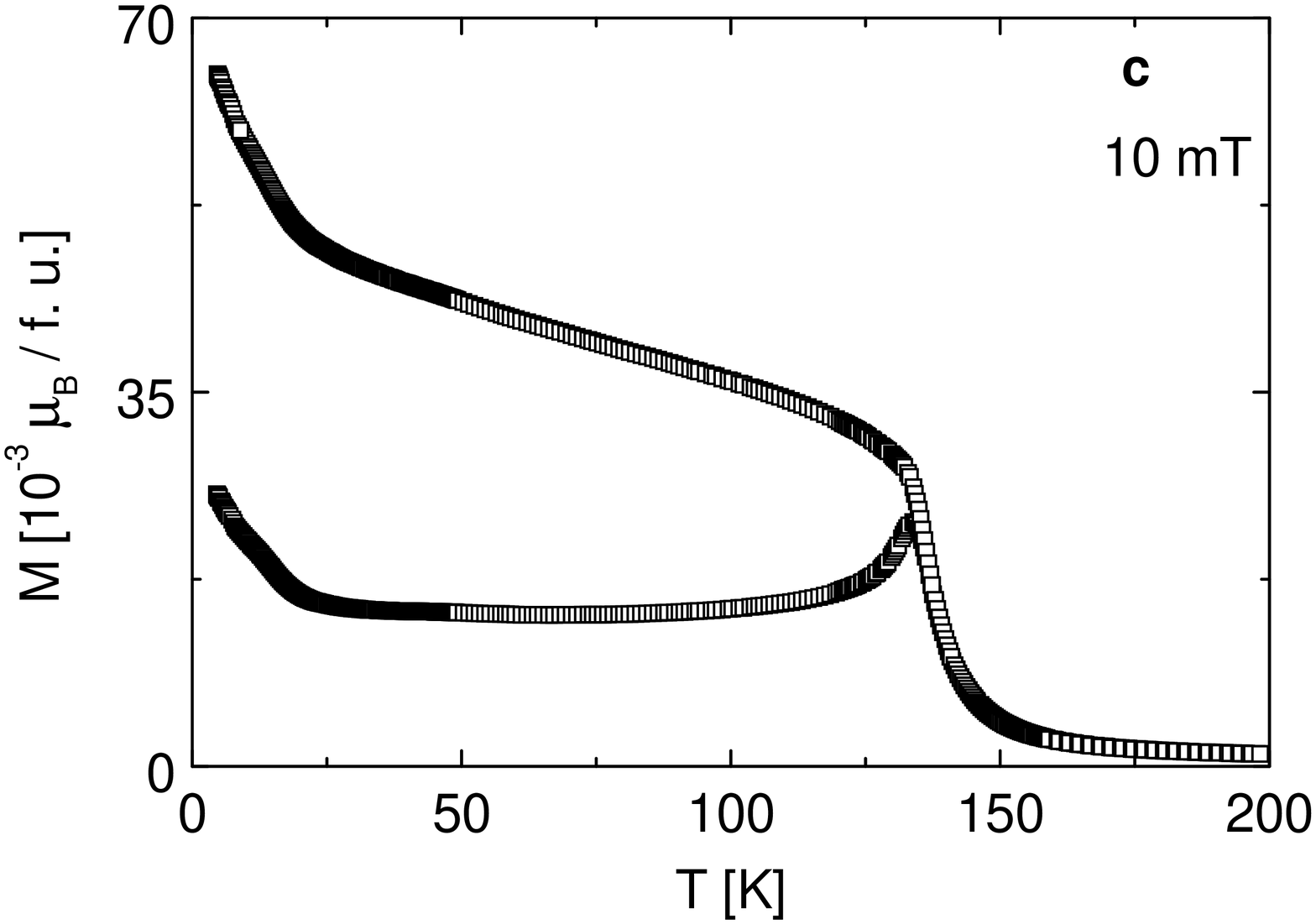}
\end{center}
\caption{Magnetic moment (in $\mu_{B}$ 
per formula unit) of our Ru-1212 sample at 0.25~mT
  (\textbf{a}), 1~mT (\textbf{b}) and 10~mT (\textbf{c}).}\label{fig:dc}
\end{figure}

The results of d.c. magnetization measurements for three different measuring
fields are shown in figure~\ref{fig:dc}.
% Begin figure3
%End figure3
At about 135~K, a magnetic transition is observed for the compound. The exact
temperature of this transition depends on the applied field and decreases
slightly from 137~K at 0.25~mT to 134~K at 10~mT. Although it was originally
believed, that this transition is due to ferromagnetic ordering of the Ru
moments, powder neutron diffraction studies \cite{Chmaissem,Lynn,Jorgensen}
revealed an antiferromagnetically ordered state of the Ru moments
and put a limit of
about 0.1~$\mu_{B}$ per Ru on any ferromagnetic component. 
The presence of a ferromagnetic component 
is indicated by hysteresis loops in M-H curves measured in the magnetically 
ordered state \cite{Bernhard2,Jorgensen}. For these results to be consistent with the neutron 
diffraction data, the antiferromagnetically ordered Ru moments must be canted to give a net 
magnetic moment. This canting is thought to be associated with rotations
of the RuO$_6$-octahedra \cite{Bauernfeind1}.
The zero field
cooled (z.f.c.)-field cooled (f.c.)  hysteresis observed for our M-T curves in
the magnetically ordered state presumably arises \cite{Felner} from an increased canting of the 
Ru moments due to the presence of the external magnetic field in the f.c. process, 
which leads to higher values of the magnetization 
compared to the z.f.c. branch. The z.f.c.--f.c. hysteresis above about
30~K is also reminiscent of spin glass behavior and could involve freezing
out of the canted component below the irreversibility temperature, which
would be defined by the point, where the z.f.c. and f.c. curves merge.

In figure~\ref{fig:dc}a, distinct anomalies of the magnetization are observed upon
entering the superconducting state. The magnetization shows a clear increase
at 25~K and reaches a peak at 15~K. In a low field (0.25~mT), the magnetization decreases
below 15~K because of the field expulsion due to superconductivity. 
While the flux expulsion does not saturate down to our lowest measuring
temperature, probably due to the small grain size
($\sim\mu\textrm{m}$) or possibly due to undetected sample imhomogeneity,
it should be noted, that the signal corresponds to a complete flux
expulsion from 25\% of the sample volume.
At 1~mT
this peak is still observed but its height has decreased and at 10~mT no
peak is present (although the curves show a cusp at low temperatures)
and the flux expulsion characteristic of a superconductor entering the Meissner state has disappeared.
Similar anomalies have also been observed by Klamut \textit{et al.}
\cite{Klamut}. Their sample had a slightly higher superconducting transition
temperature of 35~K and the peak of the magnetization was observed at a
higher temperature, above 25~K. In \cite{Klamut} two proposals were made,
without further analysis, for the interpretation of the observed anomaly in
the d.c. magnetization curve: (a) in terms of a change of the magnetic
ordering of the Ru moments upon entering the superconducting state and (b) in
terms of an anomalous flux lattice behavior.  It appears noteworthy, that such
magnetization features have not been observed in all studies of superconducting
Ru-1212 \cite{Bauernfeind1,Bernhard1,Bernhard2}. 
In our case, we take into account
the information, that Ru moments order antiferromagnetically in 2116 compounds 
\cite{Battle1,Battle2,Battle3,Battle4,Battle5,Doi} and the
fact, that we used Sr-2116 as a precursor material for the preparation of
Ru-1212. Thus, we decided to investigate the possibility, that  the
anomalies are due to  Sr-2116 impurities, which could have been present
 in our Ru-1212 samples as a
second phase in an amount $\leq$ 3 \%, in accordance with our X-ray results.

\subsection{Magnetic study of Sr$_2$GdRuO$_6$}\label{sec:resb}

\begin{figure}
\begin{center}
  \includegraphics[clip=true,width=75mm]{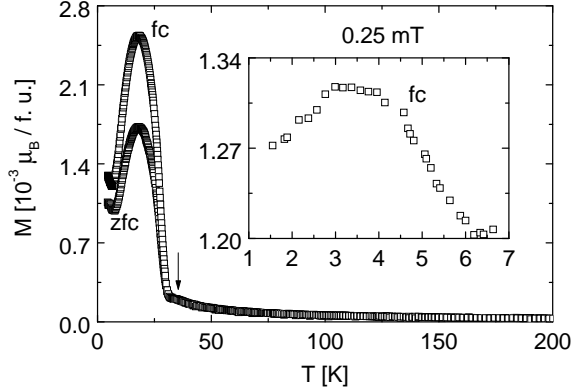}
\end{center}
\caption{Magnetic moment of Sr-2116 measured at 0.25~mT. Inset: the
  low temperature regime of the f.c. curve.}\label{fig:2116dc}
\end{figure}

\begin{figure}
\begin{center}
  \includegraphics[clip=true,width=75mm]{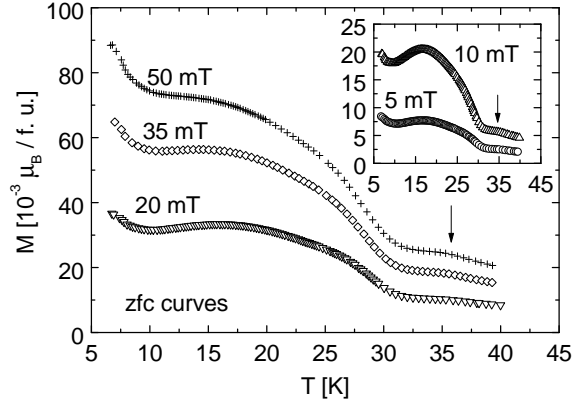}
\end{center}
\caption{The low temperature part of the z.f.c. d.c. magnetization curves of
  Sr-2116 in different magnetic fields.}\label{fig:lowtmag}
\end{figure}

\begin{figure}
\begin{center}
  \includegraphics[clip=true,width=75mm]{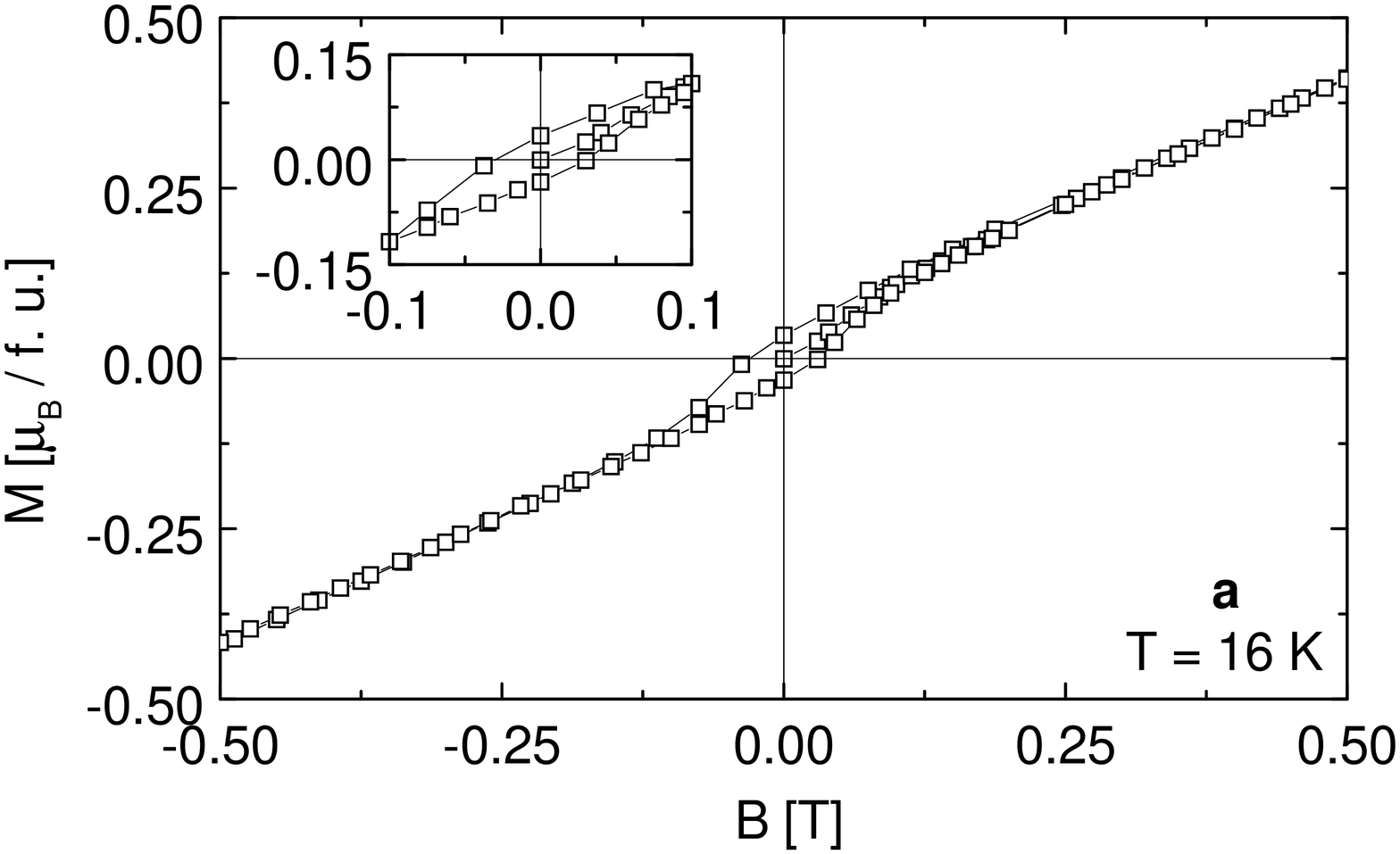}\\[2ex]
  \includegraphics[clip=true,width=75mm]{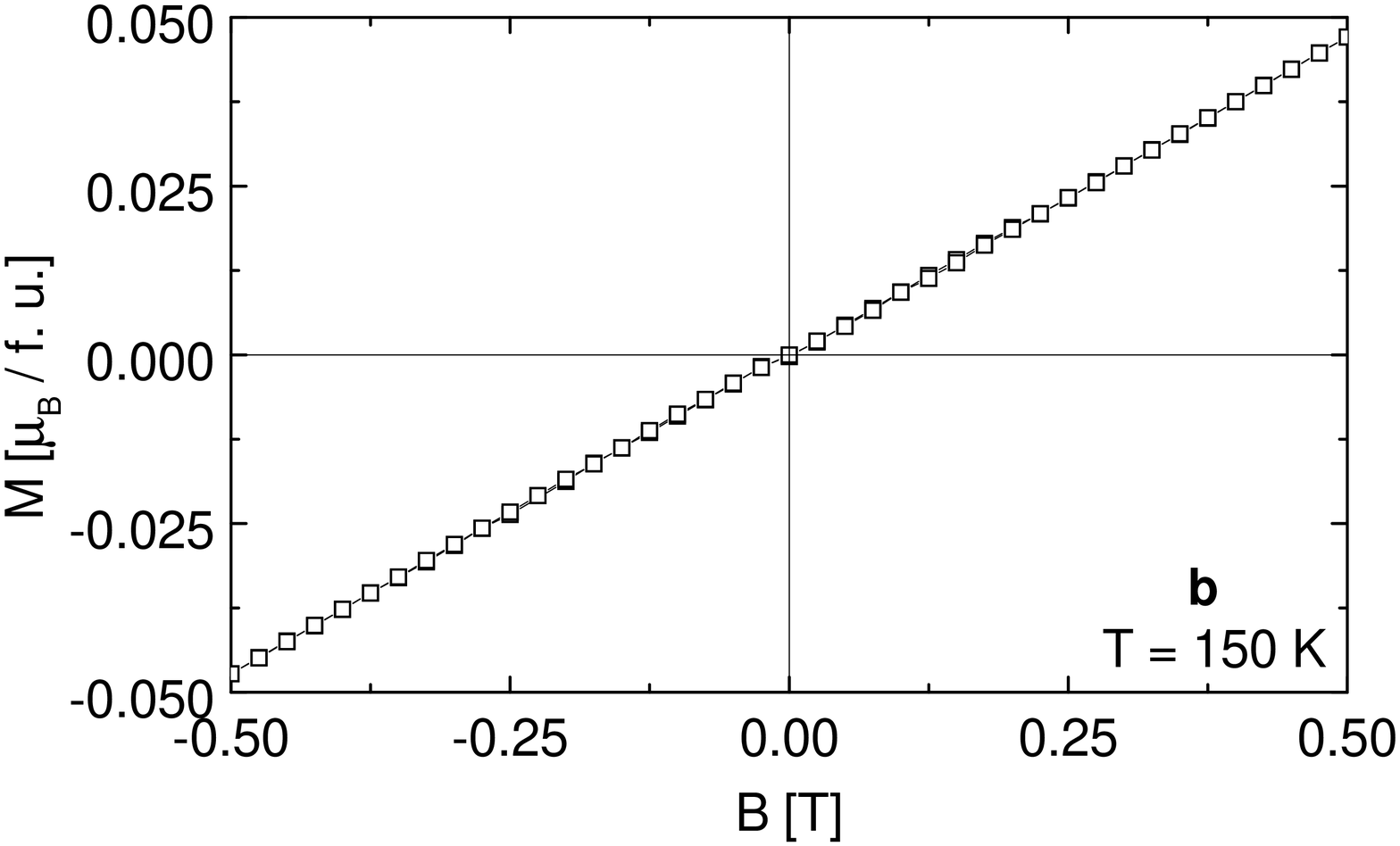}
\end{center}
\caption{d.c. magnetization measurements of Sr-2116 as a function of the magnetic field at (\textbf{a}) 16~K and (\textbf{b}) 150~K. 
Inset: the hysteresis loop in an 
  extended scale.}\label{fig:hyst}
\end{figure}

The samples of Sr-2116 were prepared as described above at the first step of the
preparation of Ru-1212. Figure~\ref{fig:2116dc} shows the z.f.c. and f.c.
magnetization taken in a field of 0.25~mT.
% Begin figure4
% End figure4
The curves consist of a high temperature paramagnetic regime, where, as
expected, the magnetization is a linear function of the magnetic field
(\textit{c.f.} figure~\ref{fig:hyst}b).
At lower temperatures, a small anomaly of the magnetization is observed between 35~K
and 30~K. The position of this cusp-like feature is marked by an arrow
in figure~\ref{fig:2116dc} and
in figure~\ref{fig:lowtmag}, where it is more clearly visible. 
From our low
temperature structure data (figure~\ref{fig:xray}b) we can exclude the
possibility, that this slight cusp is related to a structural transition.  
Below 
30~K, the magnetization begins to increase rapidly and reaches a peak
at 18~K. 
A second peak of the magnetization is observed below 3~K, as shown in the inset
of figure~\ref{fig:2116dc}.   

The temperature at which this second peak is observed is similar to the
 antiferromagnetic ordering temperature of the Gd moments in
Ru-1212, as determined by neutron diffraction \cite{Lynn} and d.c. magnetization
studies \cite{Felner}. In order to exclude the possibility, that this second
peak at 3~K or the cusp above 30~K are due to the trace impurities of cubic Gd$_2$O$_3$ present in the Sr-2116
sample, we also studied the magnetic properties of the Gd$_2$O$_3$ powder used
for sample synthesis (Chempur 99.99~\% pure). No transition was observed down
to the lowest temperature investigated (1.6~K)  in accordance with the
report by Moon and Koehler \cite{Moon}.
 
Figure~\ref{fig:lowtmag} shows, that 
the position of the cusp-like anomaly between 35~K
and 30~K as well as that of the increase of the magnetization at 30~K are
rather independent of magnetic field, while the position of the peak 
slightly shifts from 18~K at 0.25~mT down to 14~K at 35~mT and the
magnetization decrease below the peak temperature becomes less pronounced. 
% Begin figure5
% End figure5
At 50~mT, the peak has disappeared and a ferromagnetic-like curve is observed.
A f.c. magnetization curve taken at 200~mT (not shown) between 7~K and 300~K
appears paramagnetic-like throughout the whole temperature range but the cusp
in the 35-30~K temperature range is again clearly visible.

The  hysteresis between the z.f.c. and f.c. curves
(figure~\ref{fig:2116dc}) is reminiscent of that observed for Ru-1212 and
indicates, that Sr-2116 is not a simple antiferromagnet. As proposed for
Ru-1212, such a behavior could be caused by a canted arrangement
of the spins giving rise to a net magnetic moment. 
The presence of a
ferromagnetic component is indicated by the hysteresis loop obtained at 16~K, 
which is shown in figure~\ref{fig:hyst}a. Using the remanent magnetization 
as a measure of the spontaneous ferromagnetic moment this is $\sim 0.03\,\mu_{B}$ 
per formula unit. We should note, that for the measurement in figure~\ref{fig:hyst}a the field was 
changed between 6 and -6~T. For clarity only the low field part of the measurement is presented.
%Begin figure6
%End figure6

A possible scenario for the explanation of these  observations 
is the following: At about 35~K the expected \cite{Battle1,Battle2,Battle3,Battle4,Battle5} 
antiferromagnetic ordering of the Ru moments takes place. Presumably,
this corresponds to the 31~K ordering temperature of the Ru moments in Sr-2116
reported in literature \cite{Felner,Nowik}.
A canting of the Ru moments creates an internal field, 
which at about 30~K polarizes the Gd moments leading to the observed rapid
increase of the magnetization shown in figure~\ref{fig:2116dc}. 
At about 18~K, Gd-Gd antiferromagnetic interactions start to dominate over this polarizing effect 
and the magnetization decreases. The  ordering of the Gd moments takes place at about 3~K. 

This interpretation of the data is consistent with a scenario proposed in
\cite{Thompson} in order to  explain
 the magnetic properties of Gd$_2$CuO$_4$, where Cu orders antiferromagnetically at 260~K. 
Here, the canting of the Cu moments polarizes the Gd
moments at about 100~K.  Below 20~K, Gd-Gd interactions begin to
dominate, while the actual antiferromagnetic
ordering of the Gd moments, as indicated by specific heat and magnetization
measurements, occurs at 6.5~K. In an  applied magnetic field, 
the peak at 20~K is shifted to lower temperatures, 
similar to our results on Sr-2116 presented in figure~\ref{fig:lowtmag}, 
and finally  merges with the one at 6.5~K, which is associated 
to the antiferromagnetic ordering of the Gd moments. 

In a study of a series of Cu-doped 2116 compounds
\cite{Wu1,Chen,Wu2,Blackstead,Wu3} superconductivity was observed, which
appears to coexist with antiferromagnetic order of Ru$^{5+}$ ions. 
A sharp increase of the susceptibility at 30~K similar to
that in figure~\ref{fig:2116dc} was interpreted as the signature of
ferromagnetic ordering
due to double exchange between Ru ions of different valencies
\cite{Wu1,Chen,Wu2,Wu3}. 
Double exchange was introduced by Zener \cite{Zener} and in recent years
represents a standard approach for the explanation of the transport and
magnetic properties of the colossal magnetoresistance manganites, such as
La$_{1-x}$Sr$_{x}$MnO$_{3}$. There, the transfer of electrons through an
oxygen ion between Mn ions of different valencies, namely 3+ and 4+, created
through partial substitution of the trivalent La by divalent Sr, gives rise
to a ferromagnetic coupling between these ions. In the 2116 compounds it was
proposed, that partial doping with Cu, likely to be in the 3+ state, promotes
the formation of Ru$^{6+}$ ions. The double exchange mechanism would then
involve Ru$^{5+}$ and Ru$^{6+}$ ions.

In the case of Sr-2116 studied here, it is difficult to visualize the origin
of double exchange since no Cu doping is involved. On the other hand, Donohue
and McCann \cite{Donohue}, who prepared Sr-2116 compounds in a way similar to
ours, report, that their sample was oxygen deficient with a composition of
Sr$_2$GdRuO$_{5.95}$. Such a deficiency on the oxygen site could create Ru
ions of different valencies and give rise to double exchange. 
Oxygen uptake during a heat treatment in oxygen atmosphere is
then expected to affect the magnetic properties of such samples.

\begin{figure}
\begin{center}
  \includegraphics[clip=true,width=75mm]{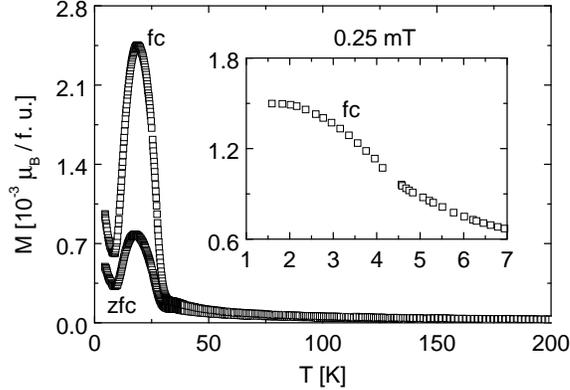}
\end{center}
\caption{z.f.c. and f.c. magnetization curves for the Sr-2116 sample of
  figure~4 after it was treated for 5~days in oxygen (see text). Inset: the
  low temperature part of the f.c. curve after the oxygen
  treatment.}\label{fig:oxy}
\end{figure}

Following these ideas, we have treated the Sr-2116 sample used for the
magnetization measurements presented in
figures~\ref{fig:2116dc}-\ref{fig:hyst} for 5~days at 1060~$^{\circ}$C in
flowing oxygen.  This treatment equals the second step in the preparation of
Ru-1212 as described above (section~\ref{sec:exp}).  In figure~\ref{fig:oxy} a
pair of M-T curves is shown, which were taken after the oxygen treatment.
% Begin figure7
% End figure7
It is obvious, that although all the features of the curves described so far
are retained and the f.c. curve remains almost unchanged,
the height of the z.f.c. peak is significantly reduced.  Incorporation of
oxygen to the sample would lead to a change of the valency of Ru ions towards
the expected one of 5+. This would then suppress double exchange. A change of
the valency of some Ru ions after the oxygen treatment would also change the
Ru-Gd interactions in the sample. This could explain why the peak shown in the
inset 
of figure~\ref{fig:2116dc}, tentatively associated with Gd-moment ordering, 
appears now (figure~\ref{fig:oxy}) to be 
shifted to a lower temperature $\leq 1.6~\textrm{K}$. The interaction of the Ru
and Gd sublattices in Ru-1212 was recently discussed in \cite{Jorgensen}.

On the basis of the above interpretation, the peak at 18~K must be related to the 
antiferromagnetic ordering of the Ru$^{5+}$ moments. This is inconsistent with the reported \cite{Felner,Nowik} 
transition temperature. Within this scenario, the 
origin of the cusp in the magnetization curve between 30 and 35~K is
unclear. It must also  
be mentioned, that the suppression of the z.f.c. magnetization seen in
figure~\ref{fig:oxy} 
could  be related to a  
change in the morphology of the sample due to the long annealing times in flowing
oxygen. 
In Ru-1212 \cite{Chmaissem} for example,  the ratio of the 
domains with opposite sense of RuO$_6$-octahedra rotation 
is assumed to depend on the annealing of the sample.

\subsection{Comparison between RuSr$_{2}$GdCu$_{2}$O$_{8}$ and
  Sr$_2$GdRuO$_{6}$}\label{sec:resc}

In view of the results presented so far it is very tempting to interpret the
anomalies of the d.c. magnetization of Ru-1212 upon entering the
superconducting state in terms of Sr-2116 impurities in the sample. 
The magnetization of Sr-2116 is similar to that of the
anomalies of Ru-1212 and the
temperatures of the magnetization peak observed for the two compounds 
differ only by about 3~K. A quantitative comparison can be done based
on the mass susceptibilities or the magnetization per gram of the
sample used for the measurements presented in
figure~\ref{fig:dc}, which may be supposed to be of unknown composition.
The z.f.c. moment of  Sr-2116 measured in a field of 0.25~mT 
(figure~\ref{fig:2116dc}) is about 
$1.7\times 10^{-3} \mu_B/\textrm{(f.u. Sr-2116)} \equiv 
1.72\times 10^{-5} \textrm{Am}^2/\textrm{g}$, 
while the peak increase above the z.f.c. ``plateau'' in the
sample of figure~\ref{fig:dc}a reaches 
$\sim 3.6\times 10^{-3} \mu_B/\textrm{(f.u. Ru-1212)} \equiv
2.9\times  10^{-5} \textrm{Am}^2/\textrm{g}$.
In order to explain the z.f.c. anomaly observed in this sample
in terms of Sr-2116 impurities one would thus need a mass 
of the impurity, which would correspond to 2.9/1.72 times the
sample mass, which, of course, is impossible. 
Different estimates are reached, if a stronger moment of the
Sr-2116 impurity is assumed, as detailed below.

On the basis of the assumption, that double exchange is causing the rapid
increase of the magnetization for Sr-2116 at 30~K one might argue, that
possible Sr-2116 impurities in the Ru-1212 sample have higher oxygen
deficiencies than the sample of figure~\ref{fig:2116dc}, which would give rise
to more pronounced magnetization peaks without requiring large amounts of
Sr-2116 impurities. One should keep in mind though, that possible impurities
underwent an oxygen treatment during the last step of the preparation of
Ru-1212. According to the result presented in figure~\ref{fig:oxy} this would
suppress double exchange.

\begin{figure}
\begin{center}
  \includegraphics[clip=true,width=75mm]{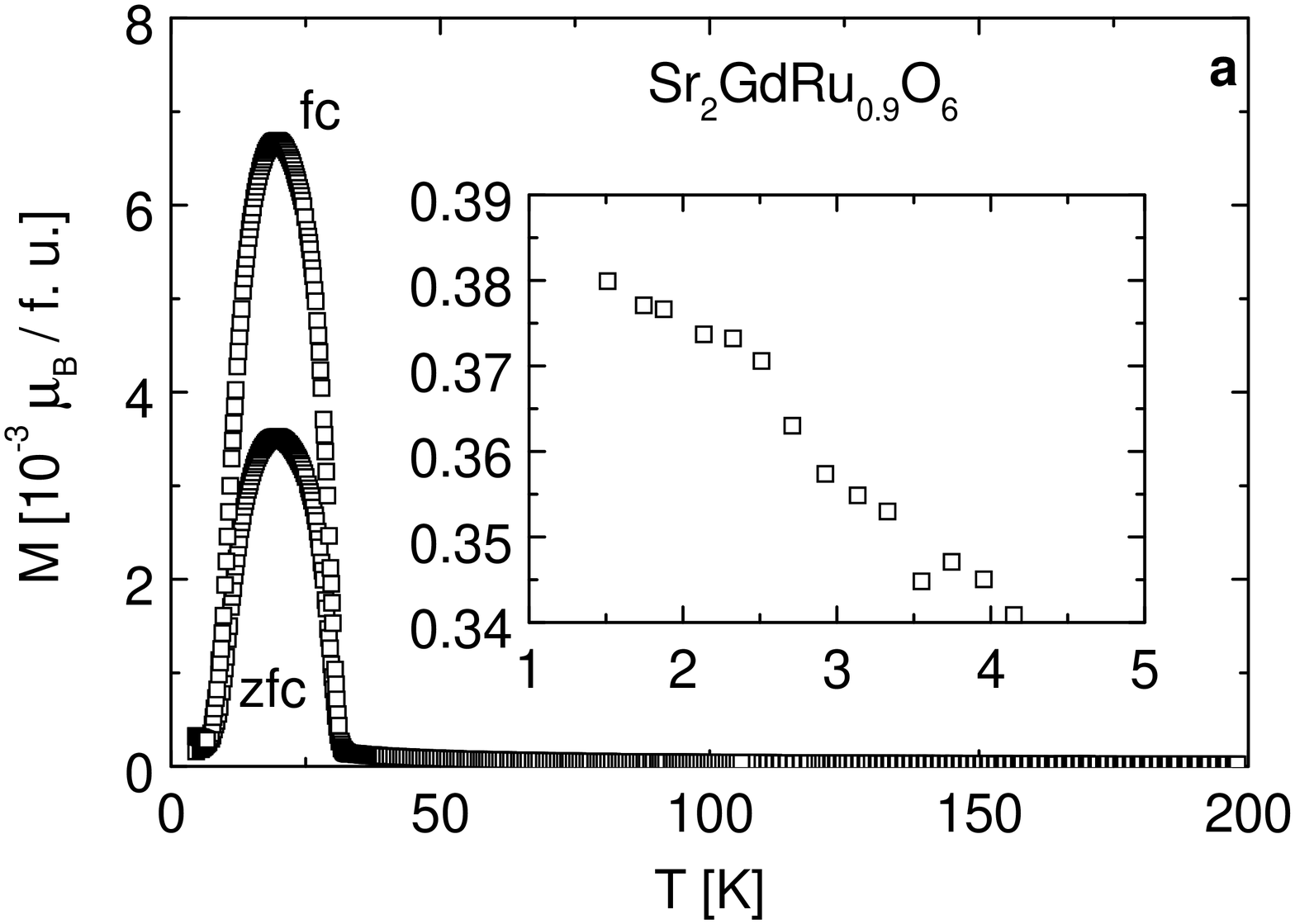}\\[2ex]
  \includegraphics[clip=true,width=75mm]{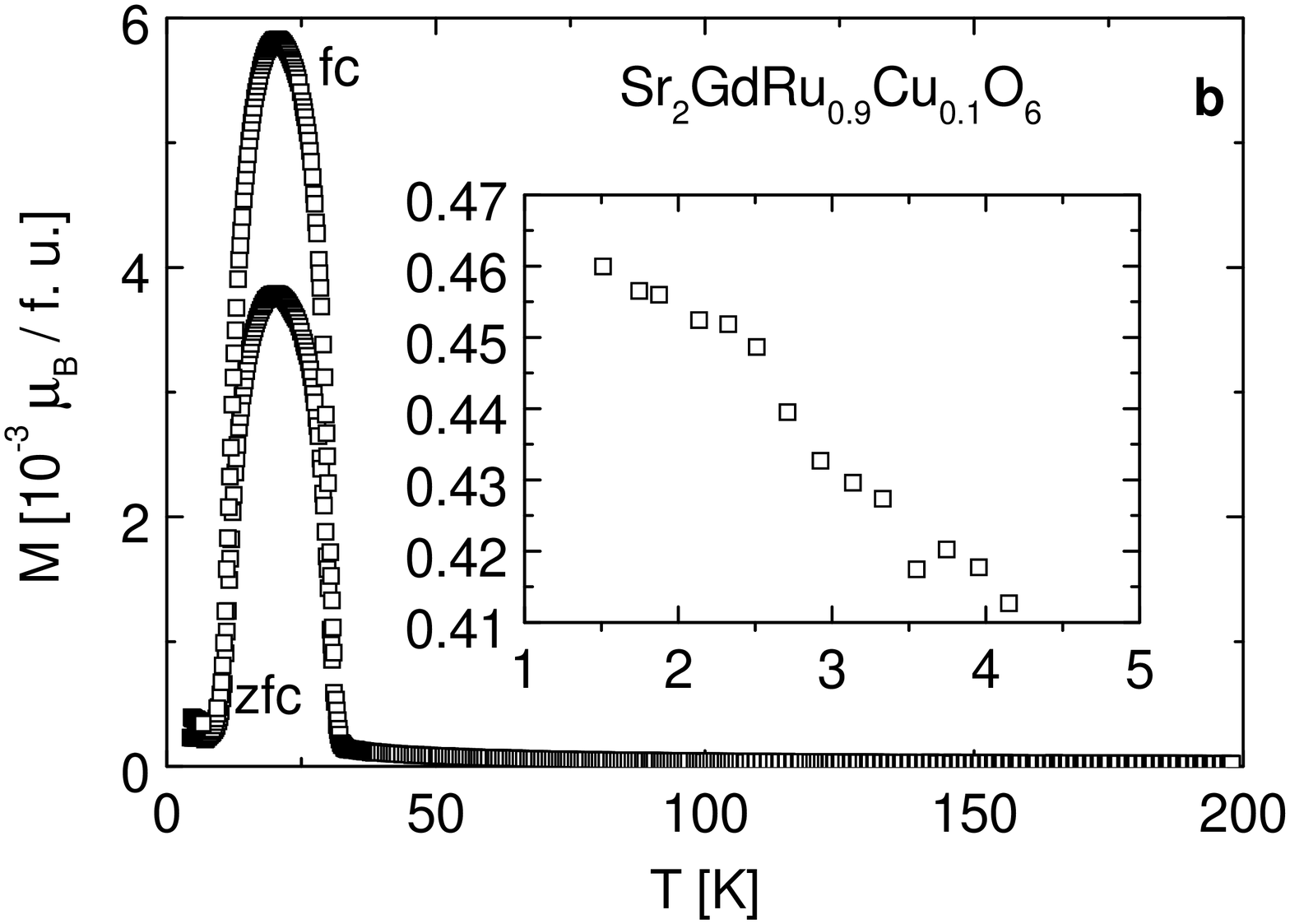}
\end{center}
\caption{d.c. magnetization curves in a measuring field of 0.25~mT of
  (\textbf{a}) Sr$_{2}$GdRu$_{0.9}$O$_{6}$ and (\textbf{b})  
  Sr$_{2}$GdRu$_{0.9}$Cu$_{0.1}$O$_{6}$ (nominal
  compositions).}\label{fig:dcdoped}
\end{figure}

Other impurities, that could be responsible for the observed magnetization
peaks in the Ru-1212 sample are Ru-deficient Sr-2116 or Cu-doped Sr-2116. 
Such samples
were also prepared and investigated with the result shown in
figure~\ref{fig:dcdoped}.
% Begin figure8
% End figure8
The sample of  Sr$_{2}$GdRu$_{0.9}$O$_{6}$  was  prepared in a way similar
to the  Sr$_{2}$GdRuO$_{6}$ sample. In order to simulate the
most likely way of formation of  Cu-doped
Sr$_{2}$GdRu$_{0.9}$Cu$_{0.1}$O$_{6}$ impurities in Ru-1212, a
Sr$_{2}$GdRu$_{0.9}$O$_{6}$ sample was used as a precursor to which CuO was
added. The pelletized mixture was heated at 1060 $^{\circ}$C for 120 h in
flowing oxygen. A comparison of the X-ray powder patterns 
of these samples (done on
the Seifert diffractometer) with that of a Sr-2116 sample is
shown in figure~\ref{fig:xray}b.  Although unidentified impurity peaks were
present (together with Gd$_{2}$O$_{3}$ trace impurities for Sr$_{2}$GdRu$_{0.9}$Cu$_{0.1}$O$_{6}$), 
we decided to investigate these
samples. The idea is, that, within the framework of double exchange for example,
 Ru deficiencies or Cu
doping could enhance the magnetism of these compounds compared to that of
Sr-2116.  
Indeed, as
shown in figure~\ref{fig:dcdoped} the magnetization of these compounds is
enhanced 
with  peaks at about 20~K, while no low
temperature peak  was observed in  the f.c. curves, which could be associated
to Gd ordering 
in these compounds. A quantitative analysis of the results, 
similar to the one above for Sr-2116, using a moment of
$4-6\times 10^{-3} \mu_B/\textrm{f.u.}$ (figure~\ref{fig:dcdoped})
leads to similar result, that a large fraction of the sample 
(50-70\%) would have to consist of the impurity phase.
Given the single-phase
nature of the x-ray powder pattern, which was taken on the sample used
for the magnetic measurements presented in figure~\ref{fig:dc}, 
this is impossible.

For the Sr$_{2}$GdRu$_{0.9}$Cu$_{0.1}$O$_{6}$ sample no indication of
superconductivity was observed in our d.c. magnetization measurements unlike
the report for other Cu doped 2116 compounds
\cite{Wu1,Chen,Wu2,Blackstead,Wu3}. It  must be mentioned, however,  that these
superconducting samples were prepared at a much higher
temperature of about 1400 $^{\circ}$C and that a similar sample with Ba in the
place of Sr, although it was prepared at high temperatures, did not show
superconductivity \cite{Blackstead}.

\section{Concluding remarks}\label{sec:summary}
In summary, we have observed distinct anomalies in the d.c. magnetization
curves  
of Ru-1212 upon entering the superconducting state.
In order to determine whether  these  
anomalies could be due to  Sr-2116 impurities in the sample,
 an extensive  
study of the magnetic properties of Sr-2116 in terms of d.c. magnetization
measurements was carried out. 
The magnetic properties of Sr-2116 are described according to the following
scenario: At $\sim$35~K   
the Ru moments order antiferromagnetically but in a canted arrangement. This
canting of the Ru moments  
creates an internal field, which polarizes the Gd moments, leading 
to a rapid increase of the  
magnetization  at about 30~K.  
At $\sim$18~K, Gd-Gd antiferromagnetic interactions
start to dominate the polarizing effect  
and the magnetization decreases. The actual antiferromagnetic ordering of
Gd takes place at about 3~K. An alternative  
(less likely) scenario for the explanation of the data based on double
exchange between Ru ions of different valencies was also presented.

In addition to (nominally) stoichiometric Sr-2116, the magnetization of
Ru-deficient and 
Cu-doped Sr-2116 compounds was also measured;  no indication for
superconductivity of the Cu doped Sr-2116 compound was found 
in d.c.
magnetization measurements.

Although similarities were observed in the magnetic behavior of Sr-2116 and
the magnetization peaks of the Ru-1212 sample, a direct quantitative
comparison clearly revealed, that 
very large amounts of Sr-2116 (or the related compounds studied), easily
detected by X-ray powder diffraction, would be required for the explanation
of the magnetization peaks in terms of impurities present in the sample. Since no
such amounts were observed, the peaks are rather an intrinsic property of
Ru-1212. 
The proposals of Klamut \textit{et al.} \cite{Klamut} for the explanation 
of this property were mentioned above in section~\ref{sec:resa}. 
The fact, that from our data and the data presented in \cite{Klamut}, the onset of the peaks occurs 
at the temperature where inter-granular coupling is established indicates, 
that the effect is related to superconductivity. 
An alternative explanation
 based on the  scenario presented above for  the magnetic 
behavior of Sr-2116 is the following: In Ru-1212,   the Ru 
moments order antiferromagnetically at $\sim$135~K in a canted arrangement, 
which creates an internal field, that in turn polarizes the Gd moments at low temperatures 
($\sim$25~K).  Below 25~K, a combination of the polarizing effect, 
the Gd-Gd antiferromagnetic interactions and the effect of superconductivity
leads to the observation of  the peak-like feature. In any case, an open 
question is why this low temperature peak in the magnetization curve of
Ru-1212  
is not observed \cite{Bauernfeind1,Bernhard1,Bernhard2} in all
d.c. magnetization studies of Ru-1212. 
A more detailed investigation of the magnetic properties of Ru-1212
samples similar to that  
presented in section~\ref{sec:resa} of this paper, which could provide
an answer to this question and clarify the origin  
of the observed peaks, is 
now in progress and details will be published in the near future.

% End of main text

% Acknowledgments

\section{Acknowledgments}\label{sec:ack}
We thank Wolfgang Ettig and Armin Dertinger for their valuable technical
support and help.

% References

% End of references

\end{document}